\documentclass[prl,twocolumn,superscriptaddress,aps,longbibliography]{revtex4-2}
\usepackage{natbib}
\usepackage{graphicx}
\usepackage{enumitem}
\DeclareGraphicsExtensions{.pdf,.png,.jpg}
\usepackage{array}
\usepackage{amsmath}
\usepackage{amsthm}
\usepackage{amssymb}    
\usepackage{latexsym}
\usepackage{amsfonts}
\usepackage{mathrsfs}
\usepackage{color}
\usepackage{bbold}
\usepackage{kotex}

\usepackage{subfigure}
\usepackage[colorlinks=true, urlcolor=blue,citecolor=blue,linkcolor=blue]{hyperref}
\usepackage{xcolor}

\begin{document}
\newcommand{\bra}[1]{\left< #1\right|}   
\newcommand{\ket}[1]{\left|#1\right>}
\newcommand{\abs}[1]{\left|#1\right|}
\newcommand{\ave}[1]{\left<#1\right>}
\newcommand{\Tr}{\mbox{Tr}}
\renewcommand{\d}[1]{\ensuremath{\operatorname{d}\!{#1}}}
\renewcommand\qedsymbol{$\blacksquare$}

\title{Metrological power of incompatible measurements}
\author{Jeongwoo Jae}
\affiliation{Department of Physics, Hanyang University, Seoul, 04763, Republic of Korea}
\author{Jiwon Lee}
\affiliation{Department of Physics, Hanyang University, Seoul, 04763, Republic of Korea}
\author{Kwang-Geol Lee}
\affiliation{Department of Physics, Hanyang University, Seoul, 04763, Republic of Korea}
\author{M. S. Kim}
\affiliation{QOLS, Blackett Laboratory, Imperial College London, London SW7 2AZ, United Kingdom}
\author{Jinhyoung Lee}
\email{hyoung@hanyang.ac.kr}
\affiliation{Department of Physics, Hanyang University, Seoul, 04763, Republic of Korea}

\begin{abstract}
We show that measurement incompatibility is a necessary resource to enhance the precision of quantum metrology. To utilize incompatible measurements, we propose a probabilistic method of operational quasiprobability (OQ) consisting of the measuring averages. OQ becomes positive semidefinite for some quantum states. We prove that Fisher information (FI), based on positive OQ, can be larger than the conventional quantum FI. Applying the proof, we show that FI of OQ can be extremely larger than quantum FI, when estimating a parameter encoded onto a qubit state with two mutually unbiased measurements. By adopting maximum likelihood estimator and linear error propagation methods, we illustrate that they achieve the high precision that our model predicts. This approach is expected to be applicable to improve quantum sensors.
\end{abstract}
\maketitle

{\em Introduction.---}Incompatibility, which tells us that two observables are not jointly measurable, is a fundamental property of quantum measurements~\cite{Busch86,Lahti2003,Heinosaari2016}. Examples are von Neumann measurements of noncommuting observables: They do not share eigenbases, so there are no measurements to specify them at once~\cite{Born1925}. Heisenberg's uncertainty relation shows this feature that uncertainties of the noncommuting measurements cannot vanish simultaneously~\cite{Robertson1929}. The incompatibility is important not only in fundamental physics, but also in quantum information science as it is a resource for nonlocality~\cite{Wolf2009,Uola2014,Quintino2014,Uola2015} and state discrimination task~\cite{Carmeli2019,Skrzypczyk2019,Uola2019}. Although its significance for the quantum information processing has been revealed, whether the incompatibility is a resource for quantum metrology~\cite{Bollinger1996,Giovannetti2004,Tsang2009,Schnabel2010,Giovannetti2011,Degen2017} remains unexplored~\cite{Guhne2023}. Recent studies show that the noncommutativity can enhance postselected metrology~\cite{Arvidsson2020,Noah2022}. However, the noncommutativity is only necessary for the measurement incompatibility; some noncommuting measurements are jointly measurable if they are generalized measurements represented by positive operator-valued measure (POVM). Here, we generalize the argument of quantum advantage in metrology using incompatibility for generalized measurements.

Fisher information (FI) is a measure to quantify metrological power of physical resources~\cite{Cramer1946,*Rao1992}. Quantum metrology estimates unknown parameters by observing a probe state with generalized measurements, where the parameters are encoded in the probe state through the interaction. By a probability model assumed for observations, an estimator suggests parameter values as estimates. FI is an upper bound of the precision that an optimal estimator can achieve, so the larger the FI is, the more precise the estimate is. Most studies have focused on quantum states of probe rather than measurements in enhancing metrology~\cite{Jonathan2008,Tan2019}, as the state of probe determines the maximum FI over all generalized measurements, called quantum FI (QFI). Moreover, incompatibility of measurements has been considered to hinder joint estimation of parameters~\cite{Liu2019} as quantum complementarity~\cite{Bohr1928} reveals trade-off relation between precisions of noncommuting observables~\cite{Gill2000,Watanabe2011,Crowley2014}.

In this work, we consider incompatible measurements to estimate a single parameter encoded in a probe, showing that the incompatibility is a resource to improve the precision of quantum metrology. To utilize incompatible measurements in metrology, we employ a probabilistic method of operational quasiprobability (OQ)~\cite{Ryu2013,*Jae2017} consisting of characteristic functions of the measurements. OQ becomes positive semidefinite for some quantum states. We prove first that FI of the positive OQ can be larger than QFI. Applying the proof, we then show that two mutually unbiased measurements can extract a large amount of information about parameter encoded onto the probe state. We perform Monte-Carlo simulations for the estimations by maximum likelihood estimator~\cite{Paris2004} and by linear error propagation~\cite{Rafal2015} methods, so to show that the precision our model predicts is attainable.

{\em Preliminaries.---}Consider $d$-outcome generalized measurements, represented by POVM $A=\{\hat{A}_a\}$ and $B=\{\hat{B}_b\}$ for outcomes $a,b \in \{0,1,\cdots,d-1\}$. POVM is a set of positive semidefinite Hermitian operators, e.g., $\hat{A}_a \ge 0$ $\forall a$, satisfying $\sum_{a=0}^{d-1}\hat{A}_a = \mathbb{1}$. $A$ and $B$ are compatible if there exists a joint measurement, represented by a POVM $J=\{ \hat{J}_{ab} \}$, such that it reproduces $A$ and $B$ as marginals,
\begin{equation}
\sum_{b=0}^{d-1} \hat{J}_{ab} = \hat{A}_a,~\text{and}~\sum_{a=0}^{d-1} \hat{J}_{ab} = \hat{B}_b,~ \forall a,b.
\end{equation}
If there is no such joint measurement, $A$ and $B$ are incompatible~\cite{Heinosaari2016,Guhne2023}.

Estimation is to make an estimate of $\theta$ which best describes $n$ observations $\{x_1,x_2,\cdots,x_n\}$, governed by a conditional probability $p(x | \theta)$ of observing outcome $x$ (or composite of outcomes, $x$). Quantum theory models the conditional probability $p(x|\hat{\varrho}_{\theta},M) = \Tr\hat{\varrho}_{\theta} \hat{M}_x$, where $\hat{\varrho}_{\theta}$ is a state of encoding parameter $\theta$ and $M=\{ \hat{M}_x \}$ is the POVM of measurement with outcomes $x$. We assume that the parameter is encoded in a state by an arbitrary quantum channel. We employ unbiased estimator $\tilde{\theta}(x)$, $\sum_x p(x|\hat{\varrho}_{\theta},M) \tilde{\theta}(x) = \theta_0$, and mean-squared-error $\Delta^2 \theta := \sum_x p(x|\hat{\varrho}_{\theta},M) (\tilde{\theta}(x) - \theta_0)^2$ to test how imprecise the estimate is. For $n$ observations, the error is bounded from below by Cram{\'e}r-Rao inequality (CRI)
\begin{eqnarray}
\label{eq:crb}
\Delta^2 \theta \ge (n{\cal{I}}_M)^{-1} \ge  (n{\cal{I}}_{\text{Q}})^{-1},
\end{eqnarray}
where FI ${\cal{I}}_M :=\sum_x p(x|\hat{\varrho}_{\theta},M) [\partial_{\theta}\log p(x|\hat{\varrho}_{\theta},M)]^2$~\cite{Cramer1946,*Rao1992} with the partial derivative $\partial_\theta$ and QFI ${\cal{I}}_{\text{Q}} := \sup_{M} {\cal{I}}_M = \Tr \hat{L}^2_{\theta} \hat{\varrho}_{\theta}$~\cite{Braunstein1994}. Here, $\hat{L}_{\theta}$ is the symmetric logarithmic derivative operator, defined by $\partial_{\theta} \hat{\varrho}_{\theta} = (\hat{L}_{\theta}\hat{\varrho}_{\theta} + \hat{\varrho}_{\theta}\hat{L}_{\theta})/2$. The second inequality in Eq.~\eqref{eq:crb}, the so-called quantum CRI (QCRI), is regarded as the ultimate lower bound of the estimation error and the measurement attaining QFI is said optimal. We remark the implication of QCRI to lead to one of our main results.

{\bf Remark.} No FI of quantum model is larger than QFI.

To assess the precision of quantum metrology, one assumes a probability model $p(x|\hat{\varrho}_{\theta},M)$ for the given probe state $\hat{\varrho}_{\theta}$ and the measurement $M$. However, there does not exist such a model for two or more incompatible measurements in quantum theory, as they are not jointly measurable. If one forces to combine incompatible measurements into a single distribution, the distribution becomes a quasiprobability, which can be negative at some points of the parameter space. Reminded that FI is defined with a regular positive probability, we employ a generalized model to incorporate incompatible measurements.

\begin{figure}[t!]
	\includegraphics[width=0.5\textwidth]{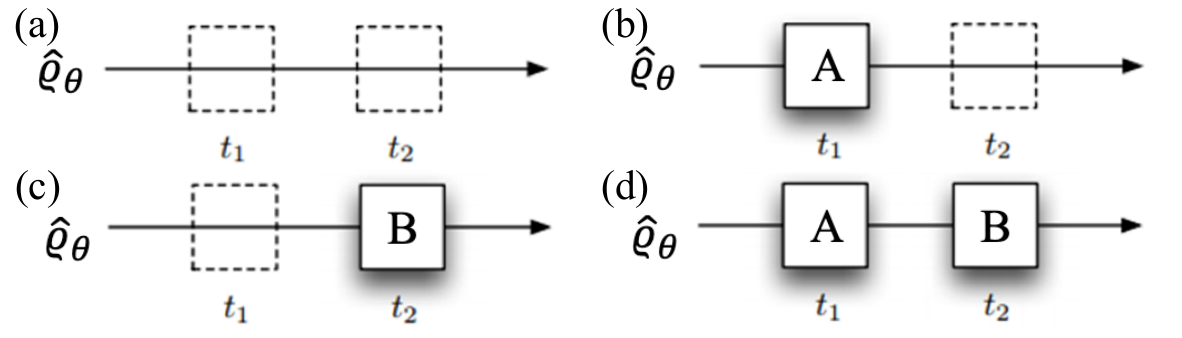}
	\caption{Four settings of measurements with the local $A$ and $B$, and their sequential measurement $S_{A\rightarrow B}$.}
\label{fig:setting}
\end{figure}

{\em Operational quasiprobability (OQ).---}Let us assume two local measurements $A = \{\hat{A}_a\}$ and $B = \{\hat{B}_b\}$, each with $d$ outcomes, and also their conjunction measurement $C = \{\hat{C}_{ab}\}$, denoted by a tuple $(A,B,C)$. As an example, for $C$, one can choose a sequential measurement $S_{A\rightarrow B}$ of performing $A$ first and $B$ later. This choice of $(A,B,S_{A\rightarrow B})$ leads to four settings of measurements depicted in Fig.~\ref{fig:setting}; (a) void measurement, (b) single local measurement $A$, (c) single local measurement $B$, and (d) sequential measurement $C=S_{A\rightarrow B}$. Then, OQ is defined by discrete Fourier transformation of the characteristic function depending on the multiple measurements, where the OQ is given by ${\cal W}(a,b) = p(a,b|C) + (p(a|A) - \sum_b p(a,b|C))/d +  ( p(b|B) - \sum_a p(a,b|C) )/d$~\cite{Ryu2013,*Jae2017}. Here, $p(\cdot|X)$ is the probability by measurement $X$. It is worth noting that $\cal W$ can be operationally constructed, for instance, by a linear optical experiment~\cite{Ryu2019}.

In quantum theory, OQ is given by ${\cal W}(a,b|\hat{\varrho}_{\theta},W)=\Tr\hat{\varrho}_{\theta}\hat{W}_{ab}$ for a probe state $\hat{\varrho}_{\theta}$. The set $W=\{\hat{W}_{ab}\}$ is a Hermitian operator-valued measure (HOVM), where
\begin{eqnarray}
\label{eq:OQ}
\hat{W}_{ab} = \hat{C}_{ab}  &+& \frac{1}{d} \left( \hat{A}_a - \sum_{b=0}^{d-1}\hat{C}_{ab} \right) +\frac{1}{d} \left( \hat{B}_b - \sum_{a=0}^{d-1}\hat{C}_{ab} \right).
\end{eqnarray}
The measure $W$ satisfies the {\em marginality},
\begin{equation}
\label{eq:marginal}
\sum_{b=0}^{d-1} \hat{W}_{ab} = \hat{A}_a,~\forall a \quad\text{and}\quad \sum_{a=0}^{d-1} \hat{W}_{ab} = \hat{B}_b,~\forall b.
\end{equation}
Note that the marginality holds for an arbitrary choice of conjunction $C$, implying OQ is a model reproducing probabilities of the local measurements by marginals. Recent work shows that (in)compatibility of the local measurements is related to positivity (negativity) of measure $W$~\cite{Jae2019}. If the two local measurements are incompatible, then $W$ should contain an element operator with at least one negative eigenvalue over any choice of $C$. On the other hand, the incompatibility is the only necessary condition for OQ to be negative-valued, as the interplay of a state and a HOVM determines the OQ. {\em Even if incompatible measurements are employed, there can exist some states for OQ to be positive semidefinite. For those states, OQ becomes a positive probability model.}

We define FI on the positive OQ and analyze how the incompatibility affects the FI. For a positive OQ, we define FI of OQ (OQFI) by
\begin{equation}
{\cal I}_{\text{OQ}} := \sum_{a,b=0}^{d-1} {\cal W}(a,b|\hat{\varrho}_{\theta}, W ) \left[{\partial_{\theta}} \log {\cal W}(a,b| \hat{\varrho}_{\theta},W )\right]^2,
\end{equation}
which determines the lower bound of mean-squared-error $\Delta^2 \theta_{\text{OQ}} := \sum_{a,b} {\cal W}(a,b| \hat{\varrho}_{\theta},W ) (\tilde{\theta}(a,b) - \theta_0)^2$, replacing FI in CRI~\eqref{eq:crb}. One of our main results is that incompatible measurements are necessary for OQFI to be larger than QFI, as stated in the following Theorem $1$.

{\bf Theorem $1$.} If there exists a state $\hat{\varrho}_{\theta}$ that OQFI is larger than QFI over any conjunction measurement $C$, the local measurements $A$ and $B$ are incompatible, i.e.,
\begin{eqnarray}
\exists \hat{\varrho}_{\theta} ~\text{s.t.}~ {\cal I}_{\text{OQ}} > {\cal I}_{\text{Q}}, \forall C ~\Rightarrow~ A~\text{and}~B~\text{are incompatible}. \nonumber
\end{eqnarray}

For two-outcome local measurements, Theorem $1$ can be rephrased. Each local measurement is represented by a noisy version of projection-valued measure (PVM) $\{\hat{P}_i\}$, i.e., a POVM $\{\mu\hat{P}_{i} + (1-\mu){\mathbb{1}}/{d}\}$, where $\hat{P}_i\ge 0$ are projectors satisfying $\Tr \hat{P}_i\hat{P}_j = \delta_{ij}$ and $\sum_{i} \hat{P}_i = \mathbb{1}$. The parameter $\mu \in [0,1]$ indicates the sharpness of measurement. The measurement is of a PVM if $\mu=1$ and, if $\mu=0$, it outputs the purely random distribution for any state.

{\bf Theorem $2$.} If there exists a {\em qubit} state $\hat{\varrho}_{\theta}$ that OQFI is larger than QFI with the sequential measurement $S_{A\rightarrow B}$, the local two-outcome measurements $A$ and $B$ are incompatible, i.e.,
\begin{eqnarray}
\exists \hat{\varrho}_{\theta}, ~ {\cal I}_{\text{OQ}} > {\cal I}_{\text{Q}} ~\text{for}~ S_{A\rightarrow B} \Rightarrow A~\text{and}~B~\text{are incompatible}. \nonumber
\end{eqnarray}

{\em Proofs of theorems.} The theorems follow from Remark and Lemmas $1$ and $2$.

{\bf Lemma $1$.} Local measurements $A$ and $B$ are compatible {\em if and only if} their HOVM $W$ is a POVM of the joint measurement for some conjunction measurement $C$.

{\bf Lemma $2$.} Local two-outcome measurements $A$ and $B$ are compatible, {\em if and only if} their HOVM $W$ is a POVM of the joint measurement with a conjunction $C=S_{A\rightarrow B}$.


See Appendices A and B for the proofs of Lemmas $1$ and $2$, respectively, and also Ref.~\cite{Jae2019} for detailed discussions.

If local measurements $A$ and $B$ are compatible, by Lemma $1$, there exists a conjunction $C$ such that $W$ is a POVM of the joint measurement $J$ for the local measurements. With the $C$, OQFI is less than or equal to QFI for all states $\hat{\varrho}_\theta$, by Remark, as OQ becomes a quantum probability model with POVM $J=W$. Thus,
\begin{eqnarray}
A~\text{and}~B~\text{are compatible} ~\Rightarrow~  \exists C ~\text{s.t.}~{\cal I}_{\text{OQ}} \le {\cal I}_{\text{Q}}, \forall \hat{\varrho}_{\theta}. \nonumber
\end{eqnarray}
The contrapositive of the synthetic proposition is Theorem $1$. Theorem $1$ becomes Theorem $2$ for a qubit with $C=S_{A\rightarrow B}$.\hfill\qedsymbol 

The theorems suggest that the incompatibility is necessary for OQ to extract more information on the parameter than any single optimal measurement. However, for a fair comparison of metrological powers between two incompatible measurements and a single optimal measurement, one needs to count the total number of measurements, while this is not accounted in the theorems.

\begin{figure}[b!]
	\includegraphics[width=0.45\textwidth]{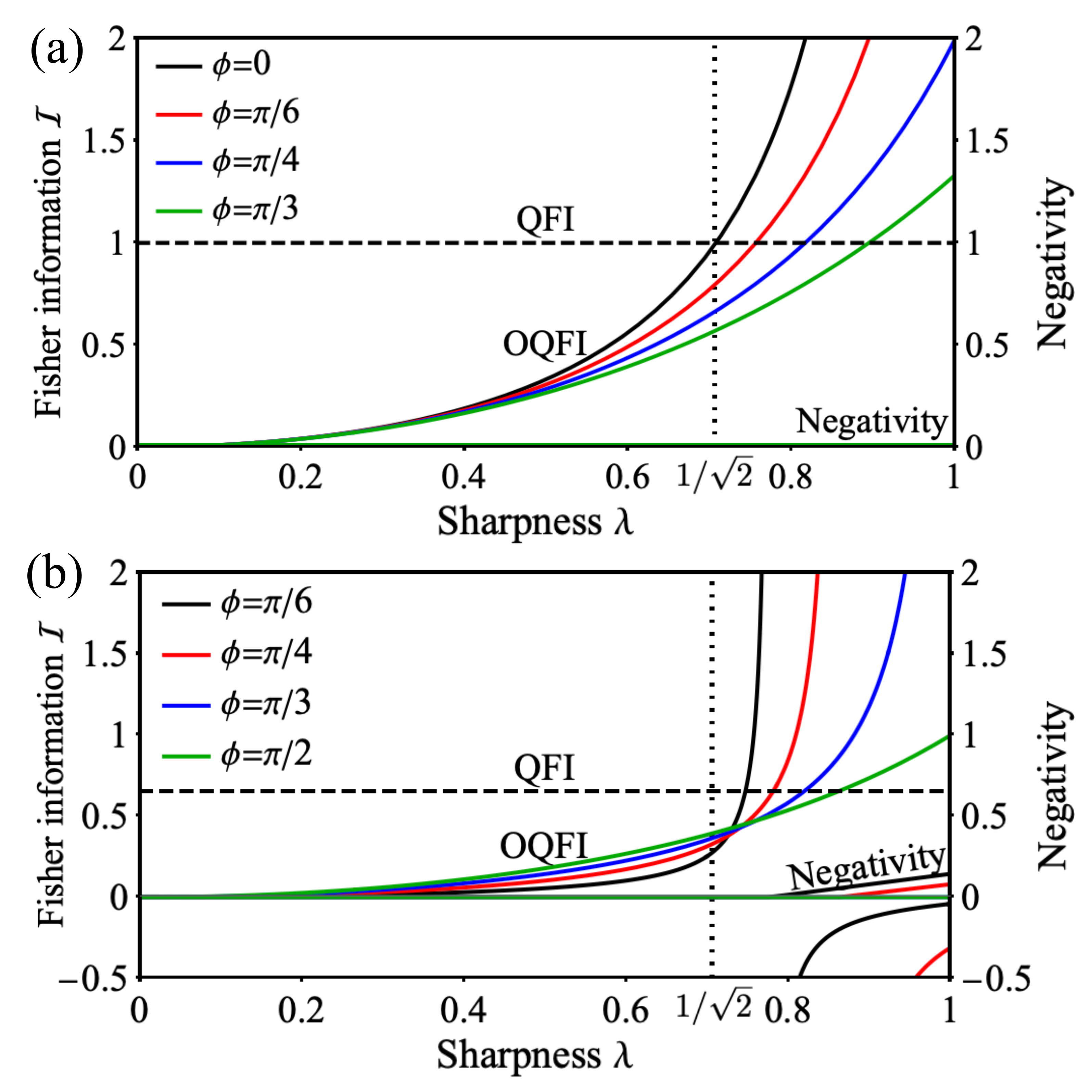}
	\caption{Comparisons of OQFIs (solid lines) and QFIs (dashed lines) in estimating (a) polar angle $\theta$ and (b) azimuthal angle $\phi$. The parameters are encoded onto qubit states as in Eq.~\eqref{eq:qubit}. The parameters are restricted to the region that OQ is positive semidefinite with the vanishing negativity ${\cal N}=0$. In (a), $\theta=\pi/2$ is to be estimated at $\phi = 0,\pi/6,\pi/4$, and $\pi/3$. Here, QFI ${\cal I}^\theta_\text{Q}=1$, regardless of $\phi$. (b) is of estimating $\phi\in\{\pi/6,\pi/4,\pi/3,\pi/2\}$ for fixed $\theta=7\pi/10$ that QFI ${\cal I}^\phi_\text{Q}=\sin^2 (7\pi/10)\approx0.654$. The vertical dotted line is of $\lambda=1/\sqrt{2}$. If their sharpness $\lambda>1/\sqrt{2}$, the local measurements are incompatible. In this region, OQFIs surpass QFIs and even diverge, depending on the parameters.}
\label{fig:FI}
\end{figure}

{\em Qubit example.---}We illustrate and compare the metrological powers with OQFI and QFI by estimating a single parameter encoded onto a qubit as an example. Pure qubit states are parameterized by
\begin{equation}\label{eq:qubit}
\ket{\psi} = \cos\left(\frac{\theta}{2}\right)\ket{0}+e^{i\phi}\sin\left(\frac{\theta}{2}\right)\ket{1},
\end{equation}
where $\{\ket{0},\ket{1}\}$ is the eigenbasis of Pauli operator $\hat{\sigma}_z$, and $\theta$ and $\phi$ are the polar and azimuthal angles in Bloch sphere. To estimate one of the parameters $\theta$ and $\phi$, we employ the tuple of measurements, $(A,B,S_{A\rightarrow B})$. Here, $A$ and $B$ are local measurements,
\begin{equation}
\label{eq:AB}
\hat{A}_{a} = \frac{1}{2} \left( {\mathbb{1}} + \omega^a \vec{\mu} \cdot \vec{\sigma}  \right)~\text{and}~\hat{B}_{b} = \frac{1}{2} \left( {\mathbb{1}} + \omega^b \vec{\nu} \cdot \vec{\sigma}  \right),
\end{equation}
where outcomes $a,b \in \{ 0, 1 \}$, and $\omega=-1$. $\vec{\sigma}=(\hat{\sigma}_x,\hat{\sigma}_y,\hat{\sigma}_z)$ are the Pauli operators. The local measurements $A$ and $B$ are assumed to be $\vec{\mu}=(0,0,\lambda)$ and $\vec{\nu}=(\lambda,0,0)$, respectively, having the same sharpness. They are mutually unbiased for nonzero $\lambda$: $\Tr \hat{A}_{a}\hat{B}_b=1/2$, $\forall~a,b$. The sequential measurement is of a POVM $S_{A\rightarrow B}=\{\hat{A}_a^{1/2} \hat{B}_b \hat{A}_a^{1/2}\}$. Then, HOVM $W$ in Eq.~\eqref{eq:OQ} is given by $W = \{ \hat{W}_{ab} = \left[ {\mathbb{1}} +\lambda\left(\omega^{a} \hat{\sigma}_z + \omega^{b}\hat{\sigma}_x\right)\right]/4\}$ and OQ is given by ${\cal W}(a,b|\psi,W)=\langle \psi | \hat{W}_{ab} | \psi\rangle$. Lemma $2$ implies that the local measurements are incompatible when their sharpness $\lambda>1/\sqrt{2}$, and then OQ can be negative. To test whether OQ is positive, we employ the {\em negativity}, defined by ${\cal N}:= \sum_{a,b} \abs{{\cal W}(a,b|\psi,W)}-1$: OQ is positive semidefinite {\em if and only if} ${\cal N}=0$.

OQFI for a parameter $g\in\{\theta,\phi\}$ reads
\begin{equation}
{\cal I}_{\text{OQ}}^g = \sum_{a,b=0}^{1} \langle \psi_g | \hat{W}_{ab} | \psi_g\rangle \left({\partial_g} \log \langle \psi_g | \hat{W}_{ab} | \psi_g\rangle \right)^2,
\end{equation}
where $\ket{\psi_g}$ indicates the dependence on the parameter $g$. We compare OQFI to QFI, given~\cite{PARIS2009} by
\begin{equation}
{\cal I}_{\text{Q}}^g = 4 \left( \langle \partial_{g}\psi | \partial_{g}\psi \rangle - \abs{\langle \psi_g | \partial_{g}\psi \rangle}^2 \right),
\end{equation}
where $\ket{\partial_{g} \psi} = \partial_{g} \ket{\psi_g}$. For estimations of $\theta$ and $\phi$, their QFIs ${\cal I}_{\text{Q}}^{\theta}=1$ and ${\cal I}_{\text{Q}}^{\phi}=\sin^2 \theta$, respectively.

Fig.~\ref{fig:FI} presents OQFIs and QFIs in estimating $\theta$ and $\phi$. For each estimation, OQFI is larger than QFI in some region of $\lambda >1/\sqrt{2}$, where the local measurements are incompatible, regardless of the type of parameter $g$. OQFI can be extremely large as $\lambda\rightarrow 1$. This result clearly shows that {\em the incompatibility of the measurements enables to extract more information than the conventional quantum estimation about the parameters.}

{\em Quantum metrological advantage.---}To compare our method to the conventional quantum estimation, the number of measurements (more precisely, the number of samples) needs to be taken into account. The comparison is made in terms of {\em advantage}
\begin{equation}
\label{eq:adv}
{\cal A}:=\log_{10}\left(\frac{{\cal I}_{\text{OQ}}}{2{\cal I}_{\text{Q}}}\right).
\end{equation}
In the qubit example, our method can employ only two settings of measurements, local measurement $B$ alone and conjunction measurement $C=S_{A\rightarrow B}$, among the four settings with $(A,B,S_{A\rightarrow B})$ in Fig.~\ref{fig:setting}. This is because the measurement $A$ alone is cancelled with the marginal of $S_{A\rightarrow B}$ by quantum theory, i.e., $\hat{A}_a-\sum_b\hat{S}_{ab} = 0$, $\forall a$ in Eq.~\eqref{eq:OQ}. When the same number of samples (i.e., quantum states) per setting are observed, the total number of samples that our method adopts is $2$ times larger than the conventional quantum estimation with a single optimal measurement. As seen in CRI~\eqref{eq:crb}, the number of $n$ samples determines the bound of CRI together with FI $\cal I$. Thus, we compare $n{\cal I}_\text{OQ}$ to $2n{\cal I}_\text{Q}$, or equivalently by ${\cal I}_\text{OQ}/2{\cal I}_\text{Q}$, basing the definition of advantage~\eqref{eq:adv}. If ${\cal I}_\text{OQ}>2{\cal I}_{\text{Q}}$, our method has the estimation error less than the lower bound of estimation error by the conventional quantum estimation, illustrating the importance of the incompatible measurements. Meanings of the advantage $\cal A$ are two folds: The local measurements are incompatible if ${{\cal A}}> \log_{10}(1/2)$ by Theorem $2$, and the incompatibility leads to the metrological advantage if ${{\cal A}}> 0$.

The advantage is attainable by maximum likelihood estimator (MLE) and linear error propagation (LEP) methods. Observing $n$ samples per setting, we obtain count numbers $c(b|B)$ and $c(a,b|S_{A\rightarrow B})$ for $B$ only and $S_{A\rightarrow B}$, respectively, where $\sum_{b} c(b|B)=n$ and $\sum_{a,b} c(a,b|S_{A\rightarrow B}) =n$. Using the numbers of counting, we construct
\begin{eqnarray}
c(a,b|W) = c(a,b|S_{A\rightarrow B})+\frac{1}{2}\left(  c(b|B) - c(b|S_{A\rightarrow B}) \right), \nonumber
\end{eqnarray}
where marginal $c(b|S_{A\rightarrow B}) = \sum_{a} c(a,b|S_{A\rightarrow B})$. (Note that, for the small number of samples, $c(a,b|W)$ can be negative by statistical fluctuations, and we omit such cases.) For each $g\in\{\theta,\phi\}$, we build a log-likelihood function
\begin{equation}
{\cal L}(g) := \frac{1}{n}\sum_{a,b=0}^{1}c(a,b|W)\log{\cal W}(a,b|\psi_{g},W).
\end{equation}
MLE estimates the parameter by ${g}_{\text{MLE}} =\operatorname*{argmax}_{g} {\cal L}(g)$. The error variance of estimates, $\Delta^2{g}_{\text{MLE}}$, approximates to $({n\cal I}_{\text{OQ}})^{-1}$ in the asymptotic limit of $n\rightarrow \infty$~\cite{lehmann2006}. We employ $(n{\cal I}_\text{obs})^{-1}$ as an estimator of error variance, where the observed Fisher information ${\cal I}_\text{obs}:=-\partial^2_g{\cal L}(g)|_{g_{\text{MLE}}}$~\cite{Bruce1997}. LEP estimates the parameter by $g_\text{LEP}$, which minimizes difference of the theoretical expectation $\langle \hat{\cal O} \rangle_g$ from a given observation $\langle{\cal O}\rangle_\text{obs}$, e.g., $g_{\text{LEP}}=\operatorname*{argmin}_{g}({\langle\hat{\cal O}\rangle_g}-\langle{\cal O}\rangle_\text{obs})^2$. Here, we employ observable $\hat{\cal O}=(-1)^{ab}\hat{W}_{ab}$, whose average is $\langle \hat{\cal O}\rangle_g = \sum_{a,b}(-1)^{ab}{\cal W}(a,b|\psi_g,W)$. By the error propagation theory~\cite{Rafal2015}, the error variance of estimates is given as
\begin{eqnarray}
\Delta^2 g_{\text{LEP}}  =\left.\frac{ \langle\Delta^2\hat{\cal O}\rangle_g}{n{(\partial_g \langle\hat{\cal O}\rangle_g )^2}}\right\vert_{g_\text{LEP}},
\end{eqnarray}
where $\langle\Delta^2\hat{\cal O}\rangle_g$ is the variance of observable $\hat{\cal O}$.


\begin{figure}[t!]
	\includegraphics[width=0.5\textwidth]{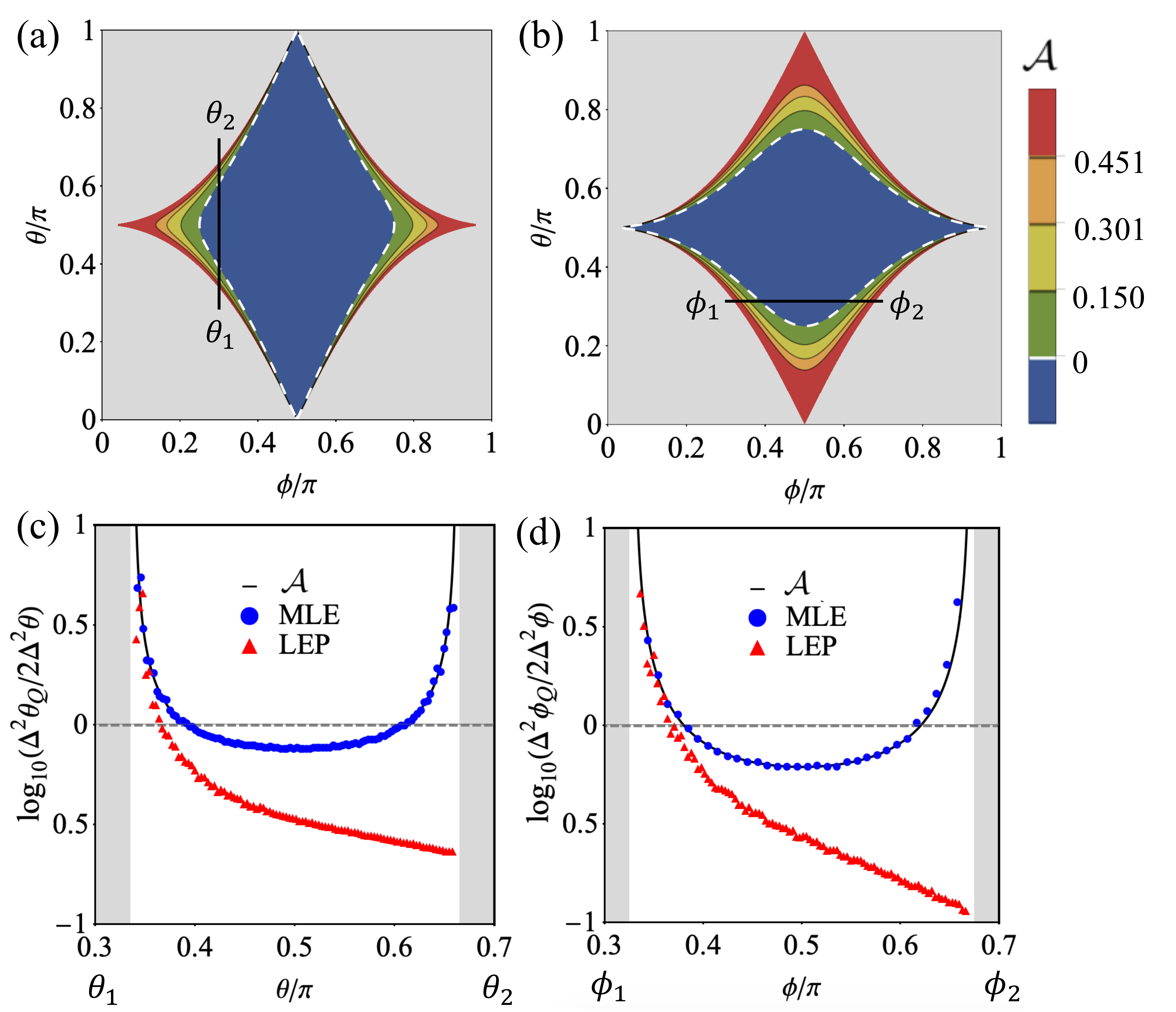}
	\caption{Metrological advantages $\cal A$ in estimation (a) $\theta$ and (b) $\phi$, employing sharp and mutually unbiased measurements. For each estimation, blue region is of no advantage with ${\cal A}<0$. Out of the region, the advantage of ${\cal A}>0$ emerges and increases extremely just before crossing to the gray region of negativity ${\cal N}>0$. Advantages for MLE and LEP to attain for (c) $\theta$ and (d) $\phi$ on line segments $\overline{\rm \theta_1 \theta_2}$ and $\overline{\rm \phi_1 \phi_2}$ in (a) and (b), respectively. The theoretic advantages $\cal A$ (black solid lines) are compared to the error-variance ratios $\log_{10}(\Delta^2 g_\text{Q}/2\Delta^2 g)$ obtained by MLE (blue circles) and LEP (red triangles), where $g\in\{\theta,\phi\}$ and $\Delta^2 g$ are the error variances of MLE and LEP for $n=10^5$ observations drawn by Monte-Carlo simulations. As references, we take $\Delta^2 g_{\text{Q}}=(n{\cal I}_{\text{Q}})^{-1}$, the theoretic lower bounds by the conventional quantum estimations.}
\label{fig:A}
\end{figure}

Fig.~\ref{fig:A} presents the theoretic advantages $\cal A$ for the parameter space $\theta, \phi \in [0,\pi]$ to estimate (a) $\theta$ and (b) $\phi$ with sharp and mutually unbiased measurements. Note that {\em the advantage $\cal A$ increases extremely just before crossing to the region of negative OQ.} It also presents the results of MLE and LEP by Monte-Carlo simulations with $n=10^5$ independent observations for each setting of measurements in estimating (c) $\theta$ and (d) $\phi$. They clearly show that the MLE attains approximately the advantage $\cal A$ over the full intervals of given parameters, and the LEP does on the some parts of intervals. 


{\em Conclusion.---}We investigate the metrological power of incompatible measurements in a single parameter estimation, employing the operational quasiprobability to incorporate the incompatible measurements in a single distribution. As a result, we prove that the incompatibility is a necessary condition to improve the precision of metrology beyond the conventional quantum estimation with quantum Fisher information (QFI). Based on the proof, we provide an estimation method with two mutually unbiased measurements on a qubit, which can achieve higher precision than the QFI limit. We perform Monte-Carlo simulations and illustrate that the high precisions are attainable by the maximum likelihood estimator and linear error propagation methods. Our scheme can be realized in a linear optical experiment~\cite{Ryu2019}, and it can also be applied to optical qubit sensors~\cite{Degen2017,Yoon2020} without entanglement.

\begin{acknowledgments}
{\em Acknowledgment.---}Authors thank to Trung Huynh, Junghee Ryu and Changhyoup Lee for discussions. KGL was supported by the National Research Foundation of Korea (NRF) grant funded by the Korea government (MSIT) (No. 2023M3K5A109481311) and Institute of Information and Communications Technology Planning \& Evaluation (IITP) grant funded by the Korea government (MSIT) (No. 2022-0-01026). MSK acknowledges the EPSRC grant (EP/T00097X/1) and AppQInfo MSCA ITN from the European Unions Horizon 2020. JL was supported by the National Research Foundation of Korea (NRF) grant funded by the Korea government (MSIT) (No. 2022M3E4A1077369).
\end{acknowledgments}

\setcounter{equation}{0}
\renewcommand{\d}[1]{\ensuremath{\operatorname{d}\!{#1}}}
\renewcommand{\thesection}{A\arabic{section}}
\renewcommand{\thesubsection}{A1-\arabic{subsection}}
\renewcommand{\theequation}{A\arabic{equation}}
\appendix


\section{Appendix A: Proof of Lemma 1}
If $W$ is a POVM of joint measurement for local measurements $A$ and $B$ and some conjunction measurement $C$, the local measurements are compatible. If the local measurements $A$ and $B$ are compatible and their joint measurement is $J$, a tuple of measurements $(A,B,J)$ results in a POVM $W$. By the marginality of $J$, the terms $\hat{A}_a - \sum_{b}\hat{J}_{ab}$ and $\hat{B}_b - \sum_{a}\hat{J}_{ab}$ in Eq.~\eqref{eq:OQ} are vanished for all outcomes $a$ and $b$, and HOVM $W$ becomes a POVM of joint measurement $J$. \hfill\qedsymbol

\section{Appendix B: Proof of Lemma 2}
The two-outcome measurements $A$ and $B$ in Eq.~\eqref{eq:AB} are compatible {\em if and only if} Busch criterion~\cite{Busch86} holds; $||\vec{\mu} + \vec{\nu}||_2 + ||\vec{\mu} - \vec{\nu}||_2 \le 2$, or equivalently $1\pm\vec{\mu}\cdot\vec{\nu}-||\vec{\mu} \pm \vec{\nu}||_2 \ge 0 $, where $||\cdot||_2$ is Euclidean norm. On the other hand, the HOVM $W$ becomes a POVM {\em if and only if} $1+\omega^{a+b}\vec{\mu}\cdot\vec{\nu} - ||\omega^{a} \vec{\mu} + \omega^{b}\vec{\nu}||_2 \ge 0$, $\forall a,b\in \{0,1\}$ as elements of HOVM $W$ for $(A,B,S_{A\rightarrow B})$ read $\hat{W}_{ab} = \left[ (1+\omega^{a+b}\vec{\mu}\cdot\vec{\nu}){\mathbb{1}} + \left(\omega^{a} \vec{\mu} + \omega^{b}\vec{\nu}\right)\cdot \vec{\sigma} \right]/4$. Thus, Busch criterion and the condition for the $W$ to be a POVM are equivalent. This implies that the local measurements are compatible {\em if and only if} HOVM $W$ of $(A,B,S_{A\rightarrow B})$ is a POVM of joint measurement. \hfill\qedsymbol


%

\end{document}